\documentclass[conference]{IEEEtran}

\IEEEoverridecommandlockouts
\usepackage{cite}
\usepackage{amsmath,amssymb,amsfonts}
\usepackage{algorithm,algorithmic}
\usepackage{graphicx}
\usepackage{textcomp}
\usepackage{xcolor}
\usepackage[acronym]{glossaries}
\usepackage{blindtext}
\def\BibTeX{{\rm B\kern-.05em{\sc i\kern-.025em b}\kern-.08em
    T\kern-.1667em\lower.7ex\hbox{E}\kern-.125emX}}
\usepackage{booktabs}
\usepackage{bm}
\usepackage{import}
\usepackage{physics}
\usepackage[hidelinks]{hyperref}
\usepackage{url}
\usepackage{multirow}
\usepackage{blindtext}
\usepackage{makecell}
\usepackage{pythonhighlight}
\usepackage{nicefrac}
\usepackage{caption}

\definecolor{color_qtm_parallel}{rgb}{0.0, 0.0, 1.0}
\definecolor{color_qtm_sequential}{rgb}{1.0, 0.5, 0.0}

\definecolor{color_qml}{rgb}{0.2, 0.2, 0.2}
\definecolor{color_qml_rev}{rgb}{0.0, 0.5, 0.0}
\definecolor{color_qml_ps}{rgb}{1.0, 0.0, 0.0}
\definecolor{color_qml_spsa}{rgb}{0.5, 0.0, 0.5}

\definecolor{modified}{rgb}{0.0, 0.0, 1.0}

\usepackage{pgfplots}
\pgfplotsset{compat=1.14}
\usepgfplotslibrary{fillbetween}

\usepackage[raggedright]{subfigure}

\usepackage[capitalize,noabbrev]{cleveref}
\Crefname{equation}{Eq.}{Eqs.}
\Crefname{figure}{Fig.}{Figs.}
\Crefname{section}{Sec.}{Secs.}
\Crefname{table}{Tab.}{Tabs.}

\newcommand\copyrighttext{%
  \footnotesize \textcopyright 2024 IEEE. Personal use of this material is permitted.
  Permission from IEEE must be obtained for all other uses, in any current or future
  media, including reprinting/republishing this material for advertising or promotional
  purposes, creating new collective works, for resale or redistribution to servers or
  lists, or reuse of any copyrighted component of this work in other works.}
\newcommand\copyrightnotice{%
\begin{tikzpicture}[remember picture,overlay]
\node[anchor=south,yshift=10pt] at (current page.south) {\fbox{\parbox{\dimexpr\textwidth-\fboxsep-\fboxrule\relax}{\copyrighttext}}};
\end{tikzpicture}%
}

\begin{document}

\newacronym{vqc}{VQC}{variational quantum circuit}
\newacronym{vqa}{VQA}{variational quantum algorithm}
\newacronym{rl}{RL}{reinforcement learning}
\newacronym{qrl}{QRL}{quantum reinforcement learning}
\newacronym{ml}{ML}{machine learning}
\newacronym{qml}{QML}{quantum machine learning}
\newacronym{dnn}{DNN}{deep neural network}
\newacronym{qnn}{QNN}{quantum neural network}
\newacronym{qc}{QC}{quantum computing}
\newacronym{nisq}{NISQ}{noisy intermediate-scale quantum}
\newacronym{pg}{PG}{policy gradient}
\newacronym{qpg}{QPG}{quantum policy gradient}
\newacronym{qnpg}{QNPG}{quantum natural policy gradient}
\newacronym{mdp}{MDP}{Markov Decision Process}
\newacronym{spsa}{SPSA}{simultaneous perturbation stochastic approximations}
\newacronym{fim}{FIM}{Fisher information matrix}
\newacronym{pdf}{PDF}{probability density function}

\title{Qiskit-Torch-Module:\\Fast Prototyping of Quantum Neural Networks
    \thanks{
    This research was conducted within the Bench-QC project, a lighthouse project of the Munich Quantum Valley initiative, which is supported by the Bavarian state with funds from the Hightech Agenda Bayern Plus.\\
    Correspondence to: nico.meyer@iis.fraunhofer.de
    }
}

\author{
    \IEEEauthorblockN{Nico Meyer\IEEEauthorrefmark{1}\IEEEauthorrefmark{2}, Christian Ufrecht\IEEEauthorrefmark{1}, Maniraman Periyasamy\IEEEauthorrefmark{1}, Axel Plinge\IEEEauthorrefmark{1}, Christopher Mutschler\IEEEauthorrefmark{1}, \\Daniel D.\ Scherer\IEEEauthorrefmark{1}, and Andreas Maier\IEEEauthorrefmark{2}
    }
    \IEEEauthorblockA{
        \IEEEauthorrefmark{1}Fraunhofer IIS, Fraunhofer Institute for Integrated Circuits IIS, Nürnberg, Germany\\
        \IEEEauthorrefmark{2}Pattern Recognition Lab, Friedrich-Alexander-Universität Erlangen-Nürnberg, Erlangen, Germany
    }
}

\maketitle

\begin{abstract}
    Quantum computer simulation software is an integral tool for the research efforts in the quantum computing community. An important aspect is the efficiency of respective frameworks, especially for training variational quantum algorithms. Focusing on the widely used \texttt{Qiskit} software environment, we develop the \texttt{qiskit-torch-module}. It improves runtime performance by two orders of magnitude over comparable libraries, while facilitating low-overhead integration with existing codebases. Moreover, the framework provides advanced tools for integrating quantum neural networks with \texttt{PyTorch}. The pipeline is tailored for single-machine compute systems, which constitute a widely employed setup in day-to-day research efforts.
\end{abstract}

\copyrightnotice  
\glsresetall
\vspace{-0.5cm}  
\section{\label{sec:intro}Introduction}

Frameworks for simulating quantum circuits have been the backbone of quantum computing research in the last decade. The collection of these tools is extensive, with the most widespread ones including \texttt{Qiskit}~\cite{Qiskit} and \texttt{Cirq}~\cite{Cirq}. Depending on system constraints, it is possible to simulate between $15$ to $25$ noise-free qubits on common consumer hardware. With access to HPC compute qubit counts in the mid $40$'s are tractable~\cite{Wu_2019,Bayraktar_2023}. If one assumes additional constraints on the quantum system, approximate tensor network methods can be used to drastically increase this number\cite{Patra_2023}. Still, for many experiments a full simulation of the state is required. Especially for prototyping new algorithms and ideas most researchers do not have access to extensive hardware clusters.

One of the most promising candidates for making use of \gls{nisq} devices are \glspl{vqa}~\cite{Preskill_2018,Cerezo_2021}. This paradigm is used in \gls{qml}~\cite{Biamonte_2017,Meyer_2022a}, where \glspl{qnn} play a central role. Other applications include quantum-enhanced optimization with QAOA~\cite{Moll_2018}, and reducing the circuit depth for chemical simulations~\cite{Grimsley_2019} with quantum computers. However, variational approaches introduce the overhead of training the underlying \glspl{vqc}, which usually is done with gradient-based optimization. To keep developing these algorithms tractable, an efficient simulation framework for training the underlying quantum models is essential.

\textbf{Related Work.} The most widely used (counting GitHub interactions) quantum computing framework \texttt{Qiskit} offers its own toolbox for \gls{qml}: \texttt{qiskit-machine-learning}~\cite{Qiskit}. Due to the prominence in the field, it is used by big parts of the \gls{qc} community to develop, test, and benchmark \glspl{vqa}. This makes performance improvements of the framework highly relevant, both for the ease of prototyping new concepts, but also with regards to e.g. energy consumption associated with the training routine. Additional frameworks for training \glspl{vqa} include \texttt{PennyLane}\cite{Bergholm_2022} and \texttt{TensorFlow Quantum}\cite{Broughton_2021}. A recent study ~\cite{Jamadagni_2024} suggests, that these toolboxes might provide performance improvements over \texttt{Qiskit}, and potentially also over \texttt{qiskit-machine-learning}. However, due to the different syntax and programming logic, re-factoring existing code between frameworks is a laborious task.

\textbf{Contribution.} We identify several shortcomings of \texttt{qiskit-machine-learning}, that limit the training efficiency of \glspl{vqa}. Our proposed alternative resolves these issues and significantly reduces runtime overhead by about two orders of magnitude. Since many \gls{qml} routines employ \glspl{qnn}, we put special attention to this setup. While \texttt{qiskit-machine-learning} allows for a basic integration of \glspl{qnn} with \texttt{PyTorch}~\cite{Paszke_2019}, we streamline this connection and enable a more advanced usage. As both, \texttt{Qiskit} and \texttt{PyTorch}, are central to our framework, we name it \texttt{qiskit-torch-module} -- short \texttt{qtm}. We emphasize that \texttt{qtm} is not intended to compete with general quantum simulation libraries. Instead, it should be viewed as a tool to speed up training \glspl{vqa} compared to \texttt{qiskit-machine-learning}. With negligible code migration overhead we observe a reduction in end-to-end computation times from hours to minutes on a representative selection of \gls{qml} algorithms. The \texttt{qtm} framework targets research efforts without access to extensive compute, where prototyping is limited to single-CPU desktop machines.

The paper is structured as follows: In \cref{sec:method}, we highlight the most important attributes of the \texttt{qiskit-torch-module}, namely simultaneous evaluation of observables, batch-parallelization, and an advanced integration with \texttt{PyTorch}. In \cref{sec:experiments} we benchmark the framework with respect to isolated metrics and on several end-to-end tasks. Finally, \cref{sec:conclusion} discusses the role of our work in the wider context of quantum computing simulators.

\section{\label{sec:method}Outline of the Module}

The proposed \texttt{qiskit-torch-module} -- from here on referred to as \texttt{qtm} -- contains several sophisticated concepts that allow for a boost of performance and usability. As indicated in the introduction, the tool should be understood as an alternative to \texttt{qiskit-machine-learning} -- abbreviated as \texttt{qml} -- for the training of \glspl{vqa}. While we have implemented several smaller alterations compared to \texttt{qml}, the main improvements constitute an efficient evaluation of multiple observables (\cref{subsec:multiobs}), batch parallelization (\cref{subsec:batchparallel}), and a straightforward integration with \texttt{PyTorch} (\cref{subsec:pytorch_integration}). Additionally, a small example of code migration is provided in~\cref{subsec:code_migration}. For further information, the reader is encouraged to refer to the documentation of the \texttt{qtm} framework.

The common objective of the \texttt{qml} and our \texttt{qtm} framework is the training of \glspl{vqa}, which incorporates quantum circuits with trainable parameters. A simple training objective can be interpreted as the cost function of a machine learning model with only one output
\begin{align}
    \label{eq:qnn_simple}
    \mathcal{M}^{\mathrm{simple}}_{\Theta,\bm{s}} &= \langle 0 | U_{\Theta,\bm{s}}^{\dagger} O U_{\Theta,\bm{s}} | 0 \rangle \\
     \label{eq:qnn_simple_obs}
    &=: \expval{O}_{\Theta,\bm{s}},
\end{align}
with some arbitrary parameterized ansatz $U_{\Theta,\bm{s}}$. Here, $\Theta$ denotes a set of trainable parameters, $\bm{s}$ refers to data encoding, and $O$ is an observable.

\subsection{\label{subsec:multiobs}Efficient Evaluation of Multiple Observables}

In practice, the envisioned quantum model is often more general than described in \cref{eq:qnn_simple,eq:qnn_simple_obs}. Typically, multiple observables are measured, which allows for an adjustable output size of the model. The generalized definition for $M$ observables reads:
\begin{align}
    \label{eq:qnn}
    \mathcal{M}_{\Theta,\bm{s}} &= \begin{bmatrix} \expval{O_0}_{\Theta,\bm{s}} \\ \vdots \\ \expval{O_{M-1}}_{\Theta,\bm{s}} \end{bmatrix}
\end{align}
If executed on actual quantum hardware, only jointly measurable observables can be evaluated in parallel. However, for simulation, this restriction does not apply. Explicitly evolving
\begin{align}
    \ket{\psi_{\Theta,\bm{s}}}:=U_{\Theta,\bm{s}} \ket{0}
\end{align}
only once allows to determine all expectation values via post-processing following $\expval{O_i}_{\Theta,\bm{s}} = \expval{\psi_{\Theta,\bm{s}} \left| O_i \right| \psi_{\Theta,\bm{s}}}$. Exploiting this simplification, the \texttt{qtm} framework achieves an approximately $M$-fold speed-up compared to \texttt{qml}, which evolves the state for each observable.

For training the model it is typically necessary to compute gradients w.r.t. the (trainable) parameters, i.e.
\begin{align}
    \label{eq:qnn_grad}
    \nabla_{\Theta}\mathcal{M}_{\Theta,\bm{s}} &= \begin{bmatrix} \nabla_{\Theta}\expval{O_0}_{\Theta,\bm{s}} \\ \vdots \\ \nabla_{\Theta}\expval{O_{M-1}}_{\Theta,\bm{s}} \end{bmatrix}.
\end{align}
A frequently employed method to compute these gradients is the parameter-shift rule~\cite{Crooks_2019}, which is also compatible with estimation on actual hardware. Unfortunately, this approach requires the simulation of $2 \cdot \abs{\Theta}$ circuits -- independent of the number of observables, if only evolving each state once as described before. This leads to long training times, even for small systems. Alternatively, it is possible to acquire an SPSA-approximation of the gradients with only $2$ circuit evaluations -- also independent of $M$. However, this technique often leads to a less stable and longer training routine~\cite{Wiedmann_2023}. If restricting to simulation, the \emph{reverse} gradient estimation technique -- also referred to as \emph{adjoint method}~\cite{Jones_2020} -- is a promising tool for small to medium-scale systems. It re-uses information from the forward pass and executes the quantum circuit in reverse, by applying adjoint operation to the intermediate states. The originally proposed method, which is also implemented in \texttt{qml}, is defined only for single-observable quantum models. To allow for efficient computation of gradients for multiple observables, we extend the original algorithm (i.e. `Algorithm 1' in~\cite{Jones_2020}). This is done in a similar manner as previously described for the expectation values. As the approach contains interleaved state evolution and evaluation of observables, the performance gain tends to be smaller compared to evaluating expectation values. However, as discussed in \cref{sec:experiments}, the improvements over \texttt{qml} are still considerable.

When using the \texttt{qtm} framework with multiple observables, these techniques are employed by default. The underlying routine for estimating expectation values is based on \texttt{qiskit.primitives.Estimator}, the gradients are computed with a modified version of \texttt{qiskit\_algorithms.ReverseEstimatorGradient}.

\begin{figure*}[t]
    \centering
    \begin{minipage}{0.47\textwidth}
    \subfigure[Combined computation times for forward and backward pass.]{
    \begin{tikzpicture}
        \centering
        \begin{axis}[
            name=plot1,
            xlabel=$\textbf{qubits}$,
            ylabel=$\textbf{time [s]}$,
            ymode=log,
            xtick={4, 6, 8, 10, 12, 14, 16, 18, 20},
            ytick={0.1, 1.0, 10.0, 100.0, 1000.0, 10000.0, 100000.0, 1000000.0, 10000000.0},
            yticklabels={, $10^0$, $10^1$, $10^2$, $10^3$, $10^4$, $10^5$, $10^6$, },
            xmin=3.9,xmax=18.1,
            ymin=0.2,ymax=5000000.0,
            label style={font=\footnotesize},
            tick label style={font=\footnotesize},
            grid=both,
            axis x line=bottom, axis y line=left,
            width=\linewidth,
            height=7.4cm,
            legend style={/tikz/every even column/.append style={column sep=0.1cm, row sep=0.1cm},at={(0.35,0.85)},anchor=east,yshift=-5mm,font=\scriptsize}]
            ]

            \addplot+[draw=none,name path=A_qtm_par,no markers] %
            	table[x=qubits,y=total_min,col sep=comma]{figures/data/qubits_qtm_parallel.csv};
            \addplot+[draw=none,name path=B_qtm_par,no markers] %
            	table[x=qubits,y=total_max,col sep=comma]{figures/data/qubits_qtm_parallel.csv};
            \addplot[color_qtm_parallel!40] fill between[of=A_qtm_par and B_qtm_par];
            \addplot[line width=.7pt,solid,color=color_qtm_parallel,mark=o] %
            	table[x=qubits,y=total,col sep=comma]{figures/data/qubits_qtm_parallel.csv};

            \addplot+[draw=none,name path=A_qtm_seq,no markers] %
            	table[x=qubits,y=total_min,col sep=comma]{figures/data/qubits_qtm_sequential.csv};
            \addplot+[draw=none,name path=B_qtm_seq,no markers] %
            	table[x=qubits,y=total_max,col sep=comma]{figures/data/qubits_qtm_sequential.csv};
            \addplot[color_qtm_sequential!40] fill between[of=A_qtm_seq and B_qtm_seq];
            \addplot[line width=.7pt,solid,color=color_qtm_sequential,mark=o] %
            	table[x=qubits,y=total,col sep=comma]{figures/data/qubits_qtm_sequential.csv};

            \addplot+[draw=none,name path=A_qml_rev,no markers] %
            	table[x=qubits,y=total_min,col sep=comma]{figures/data/qubits_qml_reverse.csv};
            \addplot+[draw=none,name path=B_qml_rev,no markers] %
            	table[x=qubits,y=total_max,col sep=comma]{figures/data/qubits_qml_reverse.csv};
            \addplot[color_qml_rev!40] fill between[of=A_qml_rev and B_qml_rev];
            \addplot[line width=.7pt,solid,color=color_qml_rev,mark=o] %
            	table[x=qubits,y=total,col sep=comma]{figures/data/qubits_qml_reverse.csv};

            \addplot+[draw=none,name path=A_qml_ps,no markers] %
            	table[x=qubits,y=total_min,col sep=comma]{figures/data/qubits_qml_ps.csv};
            \addplot+[draw=none,name path=B_qml_ps,no markers] %
            	table[x=qubits,y=total_max,col sep=comma]{figures/data/qubits_qml_ps.csv};
            \addplot[color_qml_ps!40] fill between[of=A_qml_ps and B_qml_ps];
            \addplot[line width=.7pt,solid,color=color_qml_ps,mark=o] %
            	table[x=qubits,y=total,col sep=comma]{figures/data/qubits_qml_ps.csv};

            \addplot+[draw=none,name path=A_qml_spsa,no markers] %
            	table[x=qubits,y=total_min,col sep=comma]{figures/data/qubits_qml_spsa.csv};
            \addplot+[draw=none,name path=B_qml_spsa,no markers] %
            	table[x=qubits,y=total_max,col sep=comma]{figures/data/qubits_qml_spsa.csv};
            \addplot[color_qml_spsa!40] fill between[of=A_qml_spsa and B_qml_spsa];
            \addplot[line width=.7pt,dashed,color=color_qml_spsa,mark=o,mark options={solid}] %
            	table[x=qubits,y=total,col sep=comma]{figures/data/qubits_qml_spsa.csv};

            \legend{,,,\texttt{qtm},,,,\texttt{qtm[seq]},,,,\texttt{qml-rev},,,,\texttt{qml-ps},,,,\texttt{qml-spsa}}

        \end{axis}
        \end{tikzpicture}
        }
        \end{minipage}
        \qquad
        \begin{minipage}{0.47\textwidth}
        \subfigure[Runtimes for forward pass, equivalent for all \texttt{qml} implementations.]{
        \begin{tikzpicture}
            \centering
            \begin{axis}[
                name=plot2,
                xlabel=$\textbf{qubits}$,
                ylabel=$\textbf{time [s]}$,
                ytick={0.0, 2.5, 5.0},
                yticklabels={$0$, $2.5$, $5.0$},
                xmin=3.9,xmax=18.1,
                ymin=0.0,ymax=5.5,
                restrict y to domain=0.0:25.0,
                label style={font=\footnotesize},
                tick label style={font=\footnotesize},
                grid=both,
                axis x line=bottom, axis y line=left,
                width=\linewidth,
                height=3.55cm,
                legend style={/tikz/every even column/.append style={column sep=0.1cm, row sep=0.1cm},at={(0.25,0.98)},anchor=east,yshift=-5mm,font=\scriptsize}]
                ]

                \addplot+[draw=none,name path=A_qtm_par,no markers] %
                	table[x=qubits,y=forward_min,col sep=comma]{figures/data/qubits_qtm_parallel.csv};
                \addplot+[draw=none,name path=B_qtm_par,no markers] %
                	table[x=qubits,y=forward_max,col sep=comma]{figures/data/qubits_qtm_parallel.csv};
                \addplot[color_qtm_parallel!40] fill between[of=A_qtm_par and B_qtm_par];
                \addplot[line width=.7pt,solid,color=color_qtm_parallel,mark=o] %
                	table[x=qubits,y=forward,col sep=comma]{figures/data/qubits_qtm_parallel.csv};

                \addplot+[draw=none,name path=A_qtm_seq,no markers] %
                	table[x=qubits,y=forward_min,col sep=comma]{figures/data/qubits_qtm_sequential.csv};
                \addplot+[draw=none,name path=B_qtm_seq,no markers] %
                	table[x=qubits,y=forward_max,col sep=comma]{figures/data/qubits_qtm_sequential.csv};
                \addplot[color_qtm_sequential!40] fill between[of=A_qtm_seq and B_qtm_seq];
                \addplot[line width=.7pt,solid,color=color_qtm_sequential,mark=o] %
                	table[x=qubits,y=forward,col sep=comma]{figures/data/qubits_qtm_sequential.csv};

                \addplot+[draw=none,name path=A_qml_rev,no markers] %
                	table[x=qubits,y=forward_min,col sep=comma]{figures/data/qubits_qml_reverse.csv};
                \addplot+[draw=none,name path=B_qml_rev,no markers] %
                	table[x=qubits,y=forward_max,col sep=comma]{figures/data/qubits_qml_reverse.csv};
                \addplot[color_qml!40] fill between[of=A_qml_rev and B_qml_rev];
                \addplot[line width=.7pt,solid,color=color_qml,mark=o] %
                	table[x=qubits,y=forward,col sep=comma]{figures/data/qubits_qml_reverse.csv};

                 \legend{,,,,,,,,,,,\texttt{qml}}

            \end{axis}
            \end{tikzpicture}
            }
            \\
            \subfigure[Runtimes for backward pass, \texttt{qml-spsa} only gives approximation.]{
        \begin{tikzpicture}
            \centering
            \begin{axis}[
                name=plot3,
                xlabel=$\textbf{qubits}$,
                ylabel=$\textbf{time [s]}$,
                xtick={4, 6, 8, 10, 12, 14, 16, 18, 20},
                xmin=3.9,xmax=18.1,
                ymin=0.0,ymax=110.0,
                restrict y to domain=0.0:450.0,
                label style={font=\footnotesize},
                tick label style={font=\footnotesize},
                grid=both,
                axis x line=bottom, axis y line=left,
                width=\linewidth,
                height=3.55cm,
                ]

                \addplot+[draw=none,name path=A_qtm_par,no markers] %
                	table[x=qubits,y=backward_min,col sep=comma]{figures/data/qubits_qtm_parallel.csv};
                \addplot+[draw=none,name path=B_qtm_par,no markers] %
                	table[x=qubits,y=backward_max,col sep=comma]{figures/data/qubits_qtm_parallel.csv};
                \addplot[color_qtm_parallel!40] fill between[of=A_qtm_par and B_qtm_par];
                \addplot[line width=.7pt,solid,color=color_qtm_parallel,mark=o] %
                	table[x=qubits,y=backward,col sep=comma]{figures/data/qubits_qtm_parallel.csv};

                \addplot+[draw=none,name path=A_qtm_seq,no markers] %
                	table[x=qubits,y=backward_min,col sep=comma]{figures/data/qubits_qtm_sequential.csv};
                \addplot+[draw=none,name path=B_qtm_seq,no markers] %
                	table[x=qubits,y=backward_max,col sep=comma]{figures/data/qubits_qtm_sequential.csv};
                \addplot[color_qtm_sequential!40] fill between[of=A_qtm_seq and B_qtm_seq];
                \addplot[line width=.7pt,solid,color=color_qtm_sequential,mark=o] %
                	table[x=qubits,y=backward,col sep=comma]{figures/data/qubits_qtm_sequential.csv};

                \addplot+[draw=none,name path=A_qml_rev,no markers] %
                	table[x=qubits,y=backward_min,col sep=comma]{figures/data/qubits_qml_reverse.csv};
                \addplot+[draw=none,name path=B_qml_rev,no markers] %
                	table[x=qubits,y=backward_max,col sep=comma]{figures/data/qubits_qml_reverse.csv};
                \addplot[color_qml_rev!40] fill between[of=A_qml_rev and B_qml_rev];
                \addplot[line width=.7pt,solid,color=color_qml_rev,mark=o] %
                	table[x=qubits,y=backward,col sep=comma]{figures/data/qubits_qml_reverse.csv};

                \addplot+[draw=none,name path=A_qml_ps,no markers] %
                	table[x=qubits,y=backward_min,col sep=comma]{figures/data/qubits_qml_ps.csv};
                \addplot+[draw=none,name path=B_qml_ps,no markers] %
                	table[x=qubits,y=backward_max,col sep=comma]{figures/data/qubits_qml_ps.csv};
                \addplot[color_qml_ps!40] fill between[of=A_qml_ps and B_qml_ps];
                \addplot[line width=.7pt,solid,color=color_qml_ps,mark=o] %
                	table[x=qubits,y=backward,col sep=comma]{figures/data/qubits_qml_ps.csv};

                \addplot+[draw=none,name path=A_qml_spsa,no markers] %
                	table[x=qubits,y=backward_min,col sep=comma]{figures/data/qubits_qml_spsa.csv};
                \addplot+[draw=none,name path=B_qml_spsa,no markers] %
                	table[x=qubits,y=backward_max,col sep=comma]{figures/data/qubits_qml_spsa.csv};
                \addplot[color_qml_spsa!40] fill between[of=A_qml_spsa and B_qml_spsa];
                \addplot[line width=.7pt,dashed,color=color_qml_spsa,mark=o,mark options={solid}] %
                	table[x=qubits,y=backward,col sep=comma]{figures/data/qubits_qml_spsa.csv};
    
            \end{axis}
            \end{tikzpicture}
            }
            \end{minipage}
    \captionsetup{font=footnotesize}
    \caption{\label{fig:benchmark_qubits}Benchmarking results of the proposed \texttt{qtm} module, compared to the available implementations in \texttt{qml}. All experiments are averaged over $10$ independent runs with standard deviations depicted in pale colors. The runtimes refer to a batch size of $B=48$, with single-qubit Pauli-Z observables on all qubits, and a circuit depth of $d=3$. The number of trainable parameters scales linearly in the number of qubits. (a) Depicts the combined times for one pass of computing expectation values and gradients on a logarithmic scale; (b) depicts the times for the forward pass, i.e. computing expectation values, with a cut-off at $5$ seconds; (c) depicts the times for the backward pass, i.e. computing gradients, with a cut-off at $100$ seconds.}
\end{figure*}
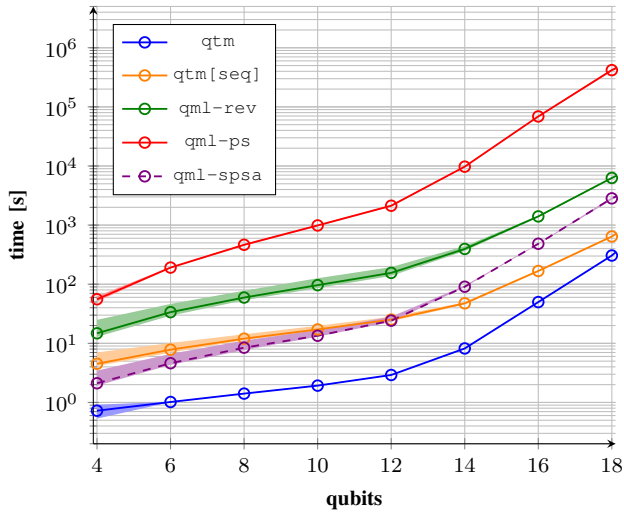
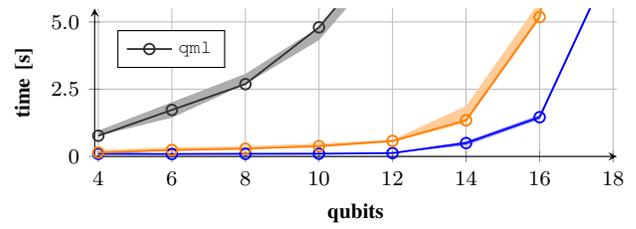
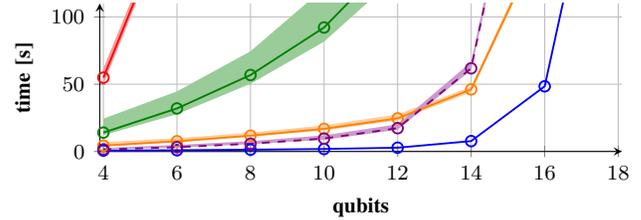

\subsection{\label{subsec:batchparallel}Parallelization for Batched Inputs}

In a typical (quantum) machine learning setting, the model is trained in a batched manner. This requires the computation of expectation values and gradients for multiple inputs $\bm{s}^{(i)},~i=0,\dots,B-1$ and a fixed set of parameter values $\Theta$:
\begin{align}
    \label{eq:qnn_batched}
    \mathcal{M}_{\Theta,\left[\bm{s}^{(i)}\right]_{i=0,\dots,B-1}} &= \begin{bmatrix} \mathcal{M}_{\Theta,\bm{s}^{(0)}} \\ \vdots \\ \mathcal{M}_{\Theta,\bm{s}^{(B-1)}} \end{bmatrix}
\end{align}
An equation for the gradients w.r.t.\ the parameters can be defined in complete correspondence to~\cref{eq:qnn_batched}. As all $B$ experiments are independent, it is possible to distribute the workload among multiple threads. If the batch size is divisible by the number of threads $T$, we equally distribute the task of handling the inputs $\bm{s}^{(i)}$ among all workers. Otherwise, we assign the first $B - \left( T \cdot \lfloor \frac{B}{T} \rfloor \right)$ threads to an additional task, to keep workloads as balanced as possible. The results of all workers are collected after completion, resulting in an array of shape $(B, M)$ for expectation values and $(B, M, \abs{\Theta})$ for gradients.

As will be demonstrated in \cref{sec:experiments}, this batch-wise parallelization technique is much more efficient than either the shot-based or circuit-based parallelization approach implemented in \texttt{qml}. Depending on the actual device, for system sizes of up to $14$ qubits this gives a close to $T$-fold speed-up. For larger systems the improvement is limited by the increasing resource requirements for state evolution, but is still noticeable.

\subsection{\label{subsec:pytorch_integration}Automatic Differentiation and Hybrid Models}

An important aspect of any software framework is its usability. For machine learning this entails automatic tracking of gradients, as realized e.g.\ by the \texttt{autograd} functionality of \texttt{PyTorch}. In fact, also the \texttt{qml} module provides such an interface via the \texttt{TorchConnector}. However, we introduce extended functionalities that are highly desirable in practice.

Many \glspl{vqa} use an underlying \gls{qnn} with multiple sets of trainable parameters, e.g. the \emph{standard} variational parameters $\bm{\theta}$ and additional input scaling parameters $\bm{\lambda}$. As both play different roles in the underlying model, typically also the optimal hyperparameters -- e.g.\ learning rates or initialization strategies -- vary. While \texttt{qml} only allows to subsume all trainable parameters in a single set with fixed hyperparameters, \texttt{qtm} allows for a fine-grained setup. We demonstrate in~\cref{subsec:endtoend} that this flexibility can indeed make a substantial difference in practice. For parameter initialization, our implementation prevents unexpected behavior caused by \texttt{Qiskit}'s value-to-parameter assignment strategy. It is always done in alphabetical order, which requires the user to guarantee consistency of \texttt{Parameter} naming with the order of provided values. The \texttt{qtm} module takes care of this pitfall and enforces the more intuitive ordering defined during model construction.

We also provide a \texttt{HybridModule}, which is end-to-end differentiable and consists of three parts: (i) a classical fully connected pre-processing layer with user-defined input size, with the output encoded in the consecutive \gls{qnn}; (ii) an instance of a \texttt{QuantumModule} with single-qubit observable measurements; (iii) a classical fully connected post-processing layer, receiving the output from the \gls{qnn}, and user-defined output size; this allows for an abstraction of problem dimensionality and qubit count, as often done in the literature.

\subsection{\label{subsec:code_migration}Migration Guide from \texttt{qml} to \texttt{qtm}}

It is straightforward to migrate implementations using \texttt{qml} to \texttt{qtm}. Let us assume an \gls{qnn}-based approach, with \texttt{VQC} denoting the underlying quantum circuit consisting of data encoding \texttt{FEATURE\_MAP} and variational layers \texttt{ANSATZ}:

\begin{python}
import qiskit_machine_learning as qml

# step 1: set up quantum neural network
qnn = qml.EstimatorQNN(circuit=VQC,
          input_params=FEATURE_MAP.parameters,
          weight_params=ANSATZ.parameters)
# step 2: connect to PyTorch, init to [0, 2*pi]
p = random.uniform(0, 2*pi, size=qnn.num_weights)
model = qml.TorchConnector(qnn, initial_weights=p)
\end{python}
As demonstrated, \texttt{qml} requires to first explicitly define a \gls{qnn}, which subsequently is connected to \texttt{Pytorch}. In \texttt{qtm}, the code snippet would be simplified as follows:
\begin{python}
import qiskit_torch_module as qtm

# set up PyTorch module and init to [0, 2*pi]
model = qtm.QuantumModule(circuit=VQC,
            encoding_params=FEATURE_MAP.parameters,
            variational_params=ANSATZ.parameters,
            variational_params_initial="uniform")
\end{python}
The \texttt{model} can subsequently be used for training and inference as any arbitrary \texttt{PyTorch} module. In more general \glspl{vqa}, the sub-modules \texttt{FastEstimator} and \texttt{FastReverseEstimatorGradient} can be used as standalone tools. As the modifications described in~\cref{subsec:multiobs,subsec:batchparallel} are realized on this level, significant speed-ups over \texttt{qiskit}'s native functionalities can be expected. The \texttt{QuantumModule} used above just provides an additional level of abstraction for convenient usage.

\section{\label{sec:experiments}Benchmarking Results}

To quantify the benefits of \texttt{qiskit-torch-module} over \texttt{qiskit-machine-learning}, we benchmark raw performance of computing expectation values and gradients in~\cref{subsec:benchmark}, backed up with end-to-end examinations in~\cref{subsec:endtoend}. We denote the proposed module as \texttt{qtm}, with \texttt{qtm[seq]} representing a sequential version without the techniques described in~\cref{subsec:batchparallel}. We compare performances to \texttt{Qiskit}'s \texttt{qml}, with \texttt{qml-rev} using reverse gradient computation, \texttt{qml-ps} resorting to the parameter-shift rule, and \texttt{qml-spsa} approximating gradients with SPSA.

We select an ansatz that is frequently employed in \gls{qml}, using data re-uploading~\cite{Perez_2020} and trainable input scaling~\cite{Jerbi_2021a}. For $n$ qubits this can be expressed as
\begin{align}
    \label{eq:concrete_ansatz}
    U_{\Theta,\bm{s}} = U_{\bm{\theta}_d,\bm{\lambda}_d,\bm{s}} \dots U_{\bm{\theta}_1,\bm{\lambda}_1,\bm{s}},
\end{align}
where $d$ denotes the depth and $\bm{s}$ is a fixed data vector. The set $\Theta := (\bm{\theta},\bm{\lambda})$ entails variational parameters $\bm{\theta}$ and state-scaling parameters $\bm{\lambda}$. The individual unitaries are realized with parameterized single-qubit rotations and a nearest-neighbor structure of CNOT-gates. The number of trainable parameters scales linearly in both the number of qubits and the circuit depth.

All experiments were conducted using \texttt{qiskit v1.0}, \texttt{qiskit-algorithms v0.3.0}, \texttt{torch v2.2.1}, \texttt{qml v0.7.1}, and \texttt{qtm v1.0}. While we recommend using the module with \texttt{qiskit}'s stable release, we also ensured backwards compatibility down to \texttt{qiskit v0.44}. All experiments in~\cref{subsec:benchmark,subsec:endtoend} were performed on a system running \texttt{Ubuntu 23.10}, with $32$ GB of RAM and a AMD Ryzen 9 5900X $12$-core CPU. In~\cref{subsec:ablation} we perform an ablation study with additional hardware arrangements and operating systems. More details are also provided in the \texttt{README} of the enclosed framework.

\subsection{\label{subsec:benchmark}Computing Expectation Values and Gradients}

The runtime needed for training a \gls{vqa} is dominated by the resources required for computing expectation values and gradients. Three features of interest are the runtime scaling with the system size, with the depth of the underlying \gls{vqc}, and the number of observables.

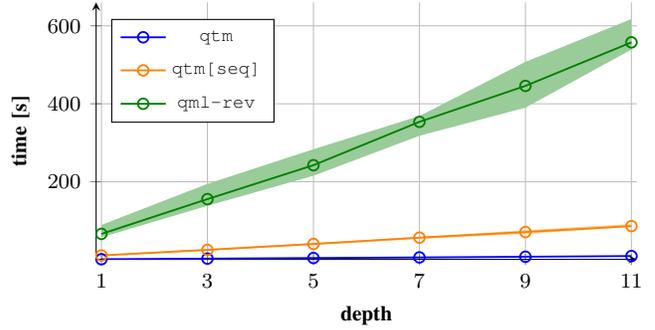
\begin{figure}[t]
    \centering

\begin{tikzpicture}
        \centering
        \begin{axis}[
            name=plot1,
            xlabel=$\textbf{depth}$,
            ylabel=$\textbf{time [s]}$,
            xtick={1, 3, 5, 7, 9, 11},
            ytick={0, 200, 400, 600},
            xmin=0.9,xmax=11.1,
            ymin=1.0,ymax=660.0,
            label style={font=\footnotesize},
            tick label style={font=\footnotesize},
            grid=both,
            axis x line=bottom, axis y line=left,
            width=0.99\linewidth,
            height=5cm,
            legend style={/tikz/every even column/.append style={column sep=0.1cm, row sep=0.1cm},at={(0.33,0.88)},anchor=east,yshift=-5mm,font=\scriptsize}]
            ]
            \addplot+[draw=none,name path=A_qtm_par,no markers] %
            	table[x=depth,y=total_min,col sep=comma]{figures/data/depth_qtm_parallel.csv};
            \addplot+[draw=none,name path=B_qtm_par,no markers] %
            	table[x=depth,y=total_max,col sep=comma]{figures/data/depth_qtm_parallel.csv};
            \addplot[color_qtm_parallel!40] fill between[of=A_qtm_par and B_qtm_par];
            \addplot[line width=.7pt,solid,color=color_qtm_parallel,mark=o] %
            	table[x=depth,y=total,col sep=comma]{figures/data/depth_qtm_parallel.csv};

            \addplot+[draw=none,name path=A_qtm_seq,no markers] %
            	table[x=depth,y=total_min,col sep=comma]{figures/data/depth_qtm_sequential.csv};
            \addplot+[draw=none,name path=B_qtm_seq,no markers] %
            	table[x=depth,y=total_max,col sep=comma]{figures/data/depth_qtm_sequential.csv};
            \addplot[color_qtm_sequential!40] fill between[of=A_qtm_seq and B_qtm_seq];
            \addplot[line width=.7pt,solid,color=color_qtm_sequential,mark=o] %
            	table[x=depth,y=total,col sep=comma]{figures/data/depth_qtm_sequential.csv};

            \addplot+[draw=none,name path=A_qml_rev,no markers] %
            	table[x=depth,y=total_min,col sep=comma]{figures/data/depth_qml_reverse.csv};
            \addplot+[draw=none,name path=B_qml_rev,no markers] %
            	table[x=depth,y=total_max,col sep=comma]{figures/data/depth_qml_reverse.csv};
            \addplot[color_qml_rev!40] fill between[of=A_qml_rev and B_qml_rev];
            \addplot[line width=.7pt,solid,color=color_qml_rev,mark=o] %
            	table[x=depth,y=total,col sep=comma]{figures/data/depth_qml_reverse.csv};

             \legend{,,,\texttt{qtm},,,,\texttt{qtm[seq]},,,,\texttt{qml-rev}}

        \end{axis}
        \end{tikzpicture}
    \captionsetup{font=footnotesize}
    \caption{\label{fig:benchmark_depth}Benchmarking results of the proposed \texttt{qtm} module, compared to \texttt{qml-rev}. The values depict gradient computation times for a batch size of $B=48$, with single-qubit Pauli-Z observables on all 12 qubits. The number of trainable parameters scales linearly in the depth. All experiments are averaged over $10$ independent runs with standard deviations depicted in pale colors. The times for \texttt{qml-ps} are too long to appear in the plot, \texttt{qml-spsa} was intentionally omitted since it only calculates approximate gradients.}
\end{figure}

In \cref{fig:benchmark_qubits}, we show the results of a runtime analysis for varying numbers of qubits. The trainable parameter count scales linearly in the system size. As expected, the resource requirements increase exponentially for all frameworks. However, \texttt{qtm} provides a speed-up of close to two orders of magnitude over \texttt{qml-rev}, and of one order of magnitude over \texttt{qml-spsa}, which only computes approximate gradients. For systems sizes over $14$ qubits the efficiency of batch-parallelization decreases slightly, since evolving the exponentially large state vector starts to dominate the total runtime. Still, even the sequential version \texttt{qtm[seq]} demonstrates a clear improvement. An additional runtime analysis of forward and backward pass reveals that the overall resources are largely consumed by gradient computation. We demonstrate in~\cref{subsec:endtoend} that the depicted absolute improvements transfer to end-to-end settings and therefore can help reduce prototyping and development overhead.

The scaling with the circuit depth -- typically proportional to the number of trainable parameters -- is shown in~\cref{fig:benchmark_depth}. The runtime scales linearly with the number of trainable parameters which is again proportional to the circuit depths. In agreement with the previous evaluations, the runtime is clearly reduced when using \texttt{qtm}. Especially when training models with increasing expressivity (related to parameter count), we expect a huge practical difference.

\begin{table}[t]
    \centering
    \begin{tabular}{@{}l|cccccccc@{}}
        \toprule
        \multirow{2}{*}{\makecell[l]{Factor\\Forward}} & \multicolumn{8}{c}{observables} \\
         & $1$  & $2$ & $4$ & $8$ & $12$ & $16$ & $24$ & $32$ \\
        \midrule
        \texttt{qtm} & $5.60$ & $11.4$ & $23.2$ & $44.4$ & $66.6$ & $81.5$ & $131$ & $172$ \\
        \texttt{qtm[seq]} & $1.17$ & $2.15$ & $4.67$ & $9.33$ & $14.0$ & $18.7$ & $25.9$ & $34.5$ \\
        \midrule
        \midrule
        \multirow{2}{*}{\makecell[l]{Factor\\Backward}} & \multicolumn{8}{c}{observables} \\
         & $1$  & $2$ & $4$ & $8$ & $12$ & $16$ & $24$ & $32$ \\
        \midrule
        \texttt{qtm} & $8.35$ & $15.6$ & $27.4$ & $43.5$ & $54.0$ & $60.1$ & $68.3$ & $74.3$ \\
        \texttt{qtm[seq]} & $1.01$ & $1.82$ & $3.11$ & $4.79$ & $5.91$ & $6.61$ & $7.47$ & $8.08$ \\
        \bottomrule
    \end{tabular}
    \vspace{0.15cm}
    \captionsetup{name=Tab.,format=hang,indention=-0.9cm,font=footnotesize}
    \caption{Improvement factors over \texttt{qml-rev} implementation for varying numbers fo measured observables. The values are averaged over $10$ independent runs of batch size $48$ and refer to $12$-qubit circuits with depth $3$, i.e. $168$ trainable parameters.}\label{tab:observables}%
\end{table}

We compare the performance of both versions of \texttt{qtm} to the best-performing implementation \texttt{qml-rev} for a increasing number of observables in~\cref{tab:observables}. This is highly relevant for e.g. multi-class classification tasks, or \gls{rl} environments with large action spaces. Additionally, it is also of interest for hybrid modules, where the output of a \gls{qnn} is fed into a classical neural network for post-processing. The results demonstrate the superiority of \texttt{qtm} in all scenarios. Indeed, for the forward pass, already \texttt{qml[seq]} improves upon \texttt{qml-rev} by a factor corresponding to the number of observables. For gradient computation, this improvement is slightly less significant, due to the complex nature of the reverse gradient estimation routine (for details see~\cref{subsec:multiobs}). Still, especially with activated batch-parallelization, we report improvements on the order of one to two magnitudes.

\subsection{\label{subsec:endtoend}Comparing End-to-End Performance}

While the isolated benchmarks in~\cref{subsec:benchmark} are helpful to anticipate the performance of the frameworks in a general setting, most important is the relevance for actual research work. In the following, we compare existing implementations using \texttt{qml} to a migrated version based on \texttt{qtm}. We want to emphasize, that the required re-factoring is typically very limited -- see also the examples provided with the framework. We selected the examples from two different sub-fields of \gls{qnn}-based \gls{qml}, namely quantum-enhanced classification and quantum reinforcement learning. We are confident that these benchmarks demonstrate the relevance of the proposed framework for a broader scope.

On of the most common applications of \glspl{qnn} is quantum-enhanced classification. Given some input data, the model is trained to predict the associated ground-truth label. A widely used benchmark dataset is MNIST~\cite{MNIST_1998}, consisting of handwritten digits with class labels from from $0$ to $9$. We select a full-quantum approach -- i.e. without trainable classical parameters. The image is encoded on $10$ qubits using incremental data-uploading~\cite{Periyasamy_2022}, interleaved with overall $220$ trainable parameters. The expectation value of single-qubit Pauli-Z observables is measured and post-processed using a softmax function. We migrate the original \texttt{qml}-based implementation to the \texttt{qtm} library -- with adjustments only being required for the setup of the quantum model.

Training is performed for $250$ epochs with a batch size of $48$ for each parameter update and a learning rate of $0.001$. The results from both implementations are in agreement with those of the original paper~\cite{Periyasamy_2022}, where the quantum model achieved an accuracy of about $55\%$ after training. Note that a random classifier would only correctly label one out of ten samples. As we are especially interested in the runtime, we compare the two realizations in~\cref{tab:times_qml}. It is evident, that using \texttt{qtm} instead of \texttt{qml} drastically reduces the time required for training the quantum classifier. The end-to-end execution time is reduced from about $14$ hours to $14$ minutes, constituting a $60$-fold improvement. This agrees with the results from ~\cref{subsec:benchmark}, where the improvement for a $10$-qubit system with $10$ observables was predicted to be in the $1.5$ to $2$ orders of magnitude range. This reduction in overall computation time enables developing refinements of the examined algorithm, as modifications can now be evaluated in a reasonable timeframe. This might allow to e.g.\ identify more sophisticated circuit ansätze that guarantee an accuracy closer to the classically achievable $98\%$.

\begin{table}[b]
    \centering
    \begin{tabular}{@{}l|ccc@{}}
        \toprule
         & total time & total steps & time per step \\
        \midrule
        \texttt{qtm} & $838$s & $12000$ & $69.8$ms \\
        \texttt{qml-rev} & $50804$s & $12000$ & $4233.7$ms \\
        \bottomrule
    \end{tabular}
    \vspace{0.15cm}
    \captionsetup{name=Tab.,format=hang,indention=-0.9cm,font=footnotesize}
    \caption{Statistics for training the full-quantum classification algorithm~\cite{Periyasamy_2022} on the MNIST dataset. The $10$-qubit \gls{vqc} incorporates $220$ trainable parameters. Pauli-Z expectation values are measured on all $10$ qubits and undergo softmax post-processing before label prediciton. Training is performed for $250$ epochs with a batch size of $48$ samples each.}\label{tab:times_qml}%
\end{table}

\Gls{qrl} has emerged as a prominent area within the field of \gls{qml}~\cite{Meyer_2022a}, with \glspl{qnn} being used as function approximators. We select a \gls{qpg} approach~\cite{Jerbi_2021a}, which is based on direct parameterization of the policy, i.e. the intended behavior. We build upon an implementation of a refined version of \gls{qpg}~\cite{Meyer_2023a}, which employs the \gls{qml} framework. This setup allows to highlight another advantage of \texttt{qtm}, namely the possibility to use different hyperparameters for multiple parameter sets. We use the ansatz employed in the original work~\cite{Meyer_2023a}, which incorporates trainable variational parameters $\bm{\theta}$ and state scaling parameters $\bm{\lambda}$. In the original implementation that utilizes the \texttt{qml} module, it is necessary to apply identical learning rates for both parameter sets.

The \texttt{CartPole-v0} experiment executed on $4$ qubits, depicted in~\cref{fig:qrl}, highlights the limitations of this rigidity. A grid search for learning rates ranging from $0.01$ to $0.1$ identified the optimal unified learning rate to be $0.05$ for both $\alpha_{\bm{\theta}}$ and $\alpha_{\bm{\lambda}}$. However, employing the \texttt{qtm} module allows to identify the distinct learning rates $\alpha_{\bm{\theta}}=0.01$ and $\alpha_{\bm{\lambda}}=0.1$. This results in significantly faster and smoother convergence. The benefits are evident in the overall training duration as presented in~\cref{tab:times_qrl}, where the time per experiment was dramatically reduced from approximately $40$ minutes to merely $2.5$ minutes. While the about $10$-fold runtime reduction for each training step is still significant, the improvement is smaller compared to the classification task. This can be attributed to the system size of only $4$ qubits and the evaluation of only a single observable. Nonetheless, both the overall computation time and also the flexibility of defining multiple hyperparameter sets highlights the benefits of \texttt{qml} over \texttt{qml}.

\begin{figure}[t]
    \centering

\begin{tikzpicture}
        \centering
        \begin{axis}[
            name=plot1,
            xlabel=$\textbf{epoch}$,
            ylabel=$\textbf{reward}$,
            ytick={0, 100, 200, 300, 400, 500},
            xtick={0, 10, 20, 30, 40},
            xmin=0,xmax=42,
            ymin=0.0,ymax=530.0,
            restrict y to domain=0.0:500.0,
            label style={font=\footnotesize},
            tick label style={font=\footnotesize},
            ymajorgrids=true,
            axis x line=bottom, axis y line=left,
            width=0.99\linewidth,
            height=4.5cm,
            legend style={/tikz/every even column/.append style={column sep=0.1cm, row sep=0.1cm},at={(0.97,0.38)},anchor=east,yshift=-5mm,font=\scriptsize}]
            ]

             \legend{,,,\texttt{qtm} ~~($\alpha_{\boldsymbol{\theta}} = 0.01\text{, }~\alpha_{\boldsymbol{\lambda}} = 0.1$),,,,\texttt{qml-rev} ~~($\alpha_{\boldsymbol{\theta}} = \alpha_{\boldsymbol{\lambda}} = 0.05$)}
            
            \addplot+[draw=none,name path=A_qtm_par,no markers] %
            	table[x=epoch,y=reward_std_upper,col sep=comma]{figures/data/qrl_qtm.csv};
            \addplot+[draw=none,name path=B_qtm_par,no markers] %
            	table[x=epoch,y=reward_std_lower,col sep=comma]{figures/data/qrl_qtm.csv};
            \addplot[color_qtm_parallel!40] fill between[of=A_qtm_par and B_qtm_par];
            \addplot[line width=.7pt,solid,color=color_qtm_parallel] %
            	table[x=epoch,y=reward_avg,col sep=comma]{figures/data/qrl_qtm.csv};

            \addplot+[draw=none,name path=A_qml_rev,no markers] %
            	table[x=epoch,y=reward_std_upper,col sep=comma]{figures/data/qrl_qml.csv};
            \addplot+[draw=none,name path=B_qml_rev,no markers] %
            	table[x=epoch,y=reward_std_lower,col sep=comma]{figures/data/qrl_qml.csv};
            \addplot[color_qml_rev!40] fill between[of=A_qml_rev and B_qml_rev];
            \addplot[line width=.7pt,solid,color=color_qml_rev] %
            	table[x=epoch,y=reward_avg,col sep=comma]{figures/data/qrl_qml.csv};

            \addplot[color=black, dashed, domain=0:100, line width=.7pt] {500.0};
            \draw[color=color_qtm_parallel, dotted, line width=1.0pt](axis cs: 13.0, 0) -- (axis cs: 13.0, 500);
            \draw[color=color_qml_rev, dotted, line width=1.0pt](axis cs: 26.0, 0) -- (axis cs: 26.0, 500);

        \end{axis}
        \end{tikzpicture}
    \captionsetup{font=footnotesize}
    \caption{\label{fig:qrl}\gls{rl} performance of the \gls{qpg} algorithm proposed in~\cite{Meyer_2023a} on \texttt{CartPole-v1} for $100$ random initializations. The performance difference originates from individual learning rates that can be set in \texttt{qtm}. In contrast, \texttt{qml} does not provide this option. Apart from this, also the runtime for each update step is reduced as depicted in~\cref{tab:times_qrl}.}
\end{figure}
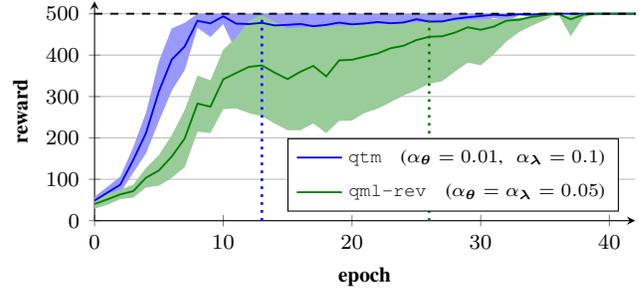
\begin{table}[tb]
    \centering
    \begin{tabular}{@{}l|ccc@{}}
        \toprule
         & total time & total steps & time per step \\
        \midrule
        \texttt{qtm} & $145.7$s & $37080$ & $3.93$ms \\
        \texttt{qml-rev} & $2301.1$s & $71259$ & $32.29$ms \\
        \bottomrule
    \end{tabular}
    \vspace{0.15cm}
    \captionsetup{name=Tab.,format=hang,indention=-0.9cm,font=footnotesize}
    \caption{Statistics for training the \gls{qpg} algorithm~\cite{Meyer_2023a} on the \texttt{CartPole-v1} environment, averaged over $10$ independent runs. The \texttt{qtm} module overall requires less steps to achieve the same performance due to flexibility of learning rates. Additionally, \texttt{qtm} also demonstrates an about $10$-fold reduction of runtime for each training step.}\label{tab:times_qrl}%
\end{table}

\subsection{\label{subsec:ablation}Ablation Study of Hardware and Operating System}

\begin{figure}[t]
    \centering
    \subfigure[\label{subfig:linux}\texttt{Ubuntu 23.10}, $64$ GB RAM, AMD Ryzen 5965WX $24$-core CPU;]{
    \begin{tikzpicture}
        \centering
        \begin{axis}[
            name=plot1,
            xlabel=$\textbf{qubits}$,
            ylabel=$\textbf{improvement}$,
            xtick={4, 6, 8, 10, 12, 14, 16, 18, 20},
            ytick={0, 50, 100, 150, 200, 250, 300},
            yticklabels={$0$, , $100$, , $200$, , $300$},
            xmin=3.9,xmax=20.2,
            ymin=0.0,ymax=310.0,
            label style={font=\footnotesize},
            tick label style={font=\footnotesize},
            grid=both,
            axis x line=bottom, axis y line=left,
            width=\linewidth,
            height=4.5cm,
            legend columns=2,
            legend style={/tikz/every even column/.append style={column sep=0.1cm, row sep=0.1cm},at={(0.68,0.97)},anchor=east,yshift=-5mm,font=\scriptsize}]
            ]

            \addplot[line width=.7pt,solid,color=color_qtm_parallel,mark=o] %
            	table[x=qubits,y=qtm_bwd,col sep=comma]{figures/data/ablation_linux.csv};

            \addplot[line width=.7pt,solid,color=color_qtm_sequential,mark=o] %
            	table[x=qubits,y=qtm_seq_bwd,col sep=comma]{figures/data/ablation_linux.csv};

            \addplot[line width=.7pt,dashed,color=color_qtm_parallel,mark=o,mark options={solid}] %
            	table[x=qubits,y=qtm_fwd,col sep=comma]{figures/data/ablation_linux.csv};

            \addplot[line width=.7pt,dashed,color=color_qtm_sequential,mark=o,mark options={solid}] %
            	table[x=qubits,y=qtm_seq_fwd,col sep=comma]{figures/data/ablation_linux.csv};

            \legend{\texttt{qtm} ~Bwd,\texttt{qtm[seq]} ~Bwd,\texttt{qtm} ~Fwd,\texttt{qtm[seq]} ~Fwd}

        \end{axis}
        \end{tikzpicture}
    }
    \\
    \subfigure[\label{subfig:windows}\texttt{Windows 10 22H2}, 16 GB RAM, Intel Core i7-10610U $4$-core CPU;]{
    \begin{tikzpicture}
        \centering
        \begin{axis}[
            name=plot2,
            xlabel=$\textbf{qubits}$,
            ylabel=$\textbf{improvement}$,
            xtick={4, 6, 8, 10, 12, 14, 16, 18},
            ytick={1, 5, 10, 15, 20},
            xmin=3.9,xmax=16.1,
            ymin=0.0,ymax=16.0,
            label style={font=\footnotesize},
            tick label style={font=\footnotesize},
            grid=both,
            axis x line=bottom, axis y line=left,
            width=\linewidth,
            height=3.2cm,
            legend style={/tikz/every even column/.append style={column sep=0.1cm, row sep=0.1cm},at={(0.35,0.85)},anchor=east,yshift=-5mm,font=\scriptsize}]
            ]

            \addplot[line width=.7pt,solid,color=color_qtm_parallel,mark=o] %
            	table[x=qubits,y=qtm_bwd,col sep=comma]{figures/data/ablation_windows.csv};

            \addplot[line width=.7pt,solid,color=color_qtm_sequential,mark=o] %
            	table[x=qubits,y=qtm_seq_bwd,col sep=comma]{figures/data/ablation_windows.csv};

            \addplot[line width=.7pt,dashed,color=color_qtm_parallel,mark=o,mark options={solid}] %
            	table[x=qubits,y=qtm_fwd,col sep=comma]{figures/data/ablation_windows.csv};

            \addplot[line width=.7pt,dashed,color=color_qtm_sequential,mark=o,mark options={solid}] %
            	table[x=qubits,y=qtm_seq_fwd,col sep=comma]{figures/data/ablation_windows.csv};


        \end{axis}
        \end{tikzpicture}
    }
    \subfigure[\label{subfig:macos}\texttt{macOS Ventura 13.6.1}, $32$ GB RAM, Intel Core i7 $6$-core CPU;]{
        \begin{tikzpicture}
        \centering
        \begin{axis}[
            name=plot3,
            xlabel=$\textbf{qubits}$,
            ylabel=$\textbf{improvement}$,
            xtick={4, 6, 8, 10, 12, 14, 16},
            ytick={1, 10, 20},
            xmin=3.9,xmax=16.1,
            ymin=0.0,ymax=22.0,
            label style={font=\footnotesize},
            tick label style={font=\footnotesize},
            grid=both,
            axis x line=bottom, axis y line=left,
            width=\linewidth,
            height=3.2cm,
            legend style={/tikz/every even column/.append style={column sep=0.1cm, row sep=0.1cm},at={(0.35,0.85)},anchor=east,yshift=-5mm,font=\scriptsize}]
            ]

            \addplot[line width=.7pt,solid,color=color_qtm_parallel,mark=o] %
            	table[x=qubits,y=qtm_bwd,col sep=comma]{figures/data/ablation_macos.csv};

            \addplot[line width=.7pt,solid,color=color_qtm_sequential,mark=o] %
            	table[x=qubits,y=qtm_seq_bwd,col sep=comma]{figures/data/ablation_macos.csv};

            \addplot[line width=.7pt,dashed,color=color_qtm_parallel,mark=o,mark options={solid}] %
            	table[x=qubits,y=qtm_fwd,col sep=comma]{figures/data/ablation_macos.csv};

            \addplot[line width=.7pt,dashed,color=color_qtm_sequential,mark=o,mark options={solid}] %
            	table[x=qubits,y=qtm_seq_fwd,col sep=comma]{figures/data/ablation_macos.csv};

        \end{axis}
        \end{tikzpicture}
    }
    \captionsetup{font=footnotesize}
    \caption{\label{fig:ablation}Ablation study testing efficiency of \texttt{qtm} on different hardware configurations and operating systems. The plots denote the runtime improvement factor over \texttt{qml-rev} on the same hardware for forward (Fwd) and backward (Bwd) pass. To ensure comparability the batch size is always selected as $4$ times the number of physical CPU cores of the respective setups.}
\end{figure}
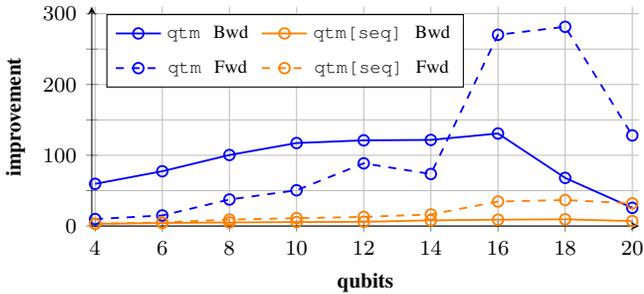
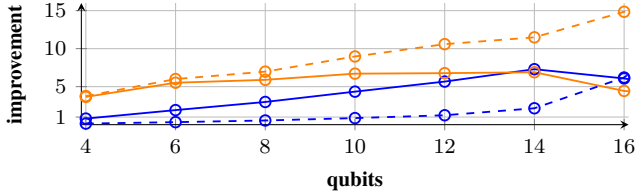
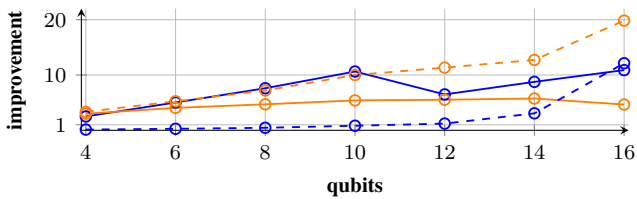

To demonstrate the efficiency of \texttt{qtm} for different hard- and software setups, we report the improvement factor compared to running the same experiment using \texttt{qml} in \cref{fig:ablation}.

The most significant speed-up is achieved on \texttt{Linux}-based operating systems as apparent from \cref{subfig:linux}, where up to $300$-fold runtime improvements are demonstrated. This is consistent with the results reported in previous sections where the optimal performance gain is given by the number of observables $M$ times the available threads $T$. Indeed, the forward pass for $16$ qubits comes close to $24 \cdot 16 = 384$. The absolute combined computation time for one input batch is reduced from $49$ minutes to $21$ seconds for this configuration.

To evaluate cross-platform compatibility we benchmarked \texttt{qtm} on notebooks with \texttt{Windows} in \cref{subfig:windows} and \texttt{MacOS} in \cref{subfig:macos}. The expected speed-up is limited by the inefficiency of Python's \texttt{multiprocessing} library on these operating systems. Overall, however, we still observe a performance improvement proportional to the number of measured observables $M$. These results underline the usefulness of our framework for various hard- and software setups.

\glsresetall

\section{\label{sec:conclusion}Discussion and Broader Scope}

We introduced the \texttt{qiskit-torch-module}, a framework for training \glspl{vqa}, with a special focus on \glspl{qnn}. It is based on a integration of \texttt{Qiskit}~\cite{Qiskit} and \texttt{PyTorch}~\cite{Paszke_2019} and resembles the structure of the respective functionalities from \texttt{qiskit-machine-learning}. The main modifications described in~\cref{sec:method} include improved measurement of multiple observables, batch parallelization, and extended functionalities for automated gradient computation. As demonstrated in~\cref{sec:experiments}, this leads to improved performance and usability of the proposed framework. This enables prototyping \gls{qml} algorithms on broadly available hardware in merely minutes.

We acknowledge that there are quantum software libraries like \texttt{PennyLane}~\cite{Bergholm_2022} or \texttt{TensorFlow Quantum}~\cite{Brockman_2016} that potentially deliver comparable performance~\cite{Jamadagni_2024}. However, within the realm of \texttt{Qiskit}-based implementations for training of \glspl{vqc}, our framework clearly improves upon the previous state of the art. It has to be noted that \texttt{qtm} was developed for classical simulation of quantum circuits~\cite{Jones_2020}. Some algorithmic improvements could be transferred to actual hardware by incorporating other primitives such as the parameter shift rule and SPSA-based approximations~\cite{Periyasamy_2024}. Given the constraints of \gls{nisq} devices and current access modalities, the classical simulation of \glspl{vqa} will remain an important part of research. The modular structure of \texttt{qtm} allows for an easy extension, e.g.\ the integration of quantum natural gradients~\cite{Stokes_2020}, which are frequently used in \gls{qml}~\cite{Meyer_2023b,Thanasilp_2023}.

Altogether, we envision \texttt{qtm} as a convenient replacement of \texttt{qml}, in the sense that almost no code re-factoring is required for existing implementations. Furthermore, the syntax closely resembles that of \texttt{qml} and therefore is easy to use for researchers already familiar with this framework. The performance improvements help to reduce prototyping times from several hours to merely minutes, drastically simplifying the task of developing and refining variational algorithms.

\section*{Acknowledgment}
We acknowledge the use and modification of parts of \texttt{Qiskit}'s code-base as a backbone of our implementation. Furthermore, the framework partially emulates structural components of \texttt{qiskit-machine-learning}, with modifications highlighted in this paper. Further details are provided in the documentation and source code of the \texttt{qiskit-torch-module}.

\section*{Code Availability and Compatibility}
The described library can be installed via \texttt{pip install qiskit-torch-module}. The code is also available at \url{https://github.com/nicomeyer96/qiskit-torch-module}. The default setup imports \texttt{qiskit v1.0} but compatibility down to \texttt{qiskit v0.44} is ensured. The enclosed \texttt{README} provides setup details and usage instructions. Further information and data is available upon reasonable request.

\newpage
\bibliographystyle{IEEEtran}
\bibliography{paper}

\begin{thebibliography}{10}
\providecommand{\url}[1]{#1}
\csname url@samestyle\endcsname
\providecommand{\newblock}{\relax}
\providecommand{\bibinfo}[2]{#2}
\providecommand{\BIBentrySTDinterwordspacing}{\spaceskip=0pt\relax}
\providecommand{\BIBentryALTinterwordstretchfactor}{4}
\providecommand{\BIBentryALTinterwordspacing}{\spaceskip=\fontdimen2\font plus
\BIBentryALTinterwordstretchfactor\fontdimen3\font minus \fontdimen4\font\relax}
\providecommand{\BIBforeignlanguage}[2]{{%
\expandafter\ifx\csname l@#1\endcsname\relax
\typeout{** WARNING: IEEEtran.bst: No hyphenation pattern has been}%
\typeout{** loaded for the language `#1'. Using the pattern for}%
\typeout{** the default language instead.}%
\else
\language=\csname l@#1\endcsname
\fi
#2}}
\providecommand{\BIBdecl}{\relax}
\BIBdecl

\bibitem{Qiskit}
{Qiskit contributors}, ``Qiskit: An open-source framework for quantum computing,'' 2023.

\bibitem{Cirq}
{Cirq developers}, ``Cirq: An open source framework for programming quantum computers,'' 2023.

\bibitem{Wu_2019}
X.-C. Wu, S.~Di, E.~M. Dasgupta, F.~Cappello, H.~Finkel, Y.~Alexeev, and F.~T. Chong, ``Full-state quantum circuit simulation by using data compression,'' in \emph{International Conference for High Performance Computing, Networking, Storage and Analysis}, 2019, pp. 1--24.

\bibitem{Bayraktar_2023}
H.~Bayraktar, A.~Charara, D.~Clark, S.~Cohen, T.~Costa, Y.-L.~L. Fang, Y.~Gao, J.~Guan, J.~Gunnels, A.~Haidar \emph{et~al.}, ``cuquantum sdk: a high-performance library for accelerating quantum science,'' in \emph{2023 IEEE International Conference on Quantum Computing and Engineering (QCE)}, vol.~1, 2023, pp. 1050--1061.

\bibitem{Patra_2023}
S.~Patra, S.~S. Jahromi, S.~Singh, and R.~Orus, ``Efficient tensor network simulation of {IBM}'s largest quantum processors,'' \emph{arXiv:2309.15642}, 2023.

\bibitem{Preskill_2018}
J.~Preskill, ``Quantum computing in the nisq era and beyond,'' \emph{Quantum}, vol.~2, p.~79, 2018.

\bibitem{Cerezo_2021}
M.~Cerezo, A.~Arrasmith, R.~Babbush, S.~C. Benjamin, S.~Endo, K.~Fujii, J.~R. McClean, K.~Mitarai, X.~Yuan, L.~Cincio \emph{et~al.}, ``Variational quantum algorithms,'' \emph{Nat. Rev. Phys.}, vol.~3, no.~9, pp. 625--644, 2021.

\bibitem{Biamonte_2017}
J.~Biamonte, P.~Wittek, N.~Pancotti, P.~Rebentrost, N.~Wiebe, and S.~Lloyd, ``Quantum machine learning,'' \emph{Nature}, vol. 549, no. 7671, pp. 195--202, 2017.

\bibitem{Meyer_2022a}
N.~Meyer, C.~Ufrecht, M.~Periyasamy, D.~D. Scherer, A.~Plinge, and C.~Mutschler, ``{A} {S}urvey on {Q}uantum {R}einforcement {L}earning,'' \emph{arXiv:2211.03464}, 2022.

\bibitem{Moll_2018}
N.~Moll, P.~Barkoutsos, L.~S. Bishop, J.~M. Chow, A.~Cross, D.~J. Egger, S.~Filipp, A.~Fuhrer, J.~M. Gambetta, M.~Ganzhorn \emph{et~al.}, ``Quantum optimization using variational algorithms on near-term quantum devices,'' \emph{Quantum Sci. Technol.}, vol.~3, no.~3, p. 030503, 2018.

\bibitem{Grimsley_2019}
H.~R. Grimsley, S.~E. Economou, E.~Barnes, and N.~J. Mayhall, ``An adaptive variational algorithm for exact molecular simulations on a quantum computer,'' \emph{Nat. Commun.}, vol.~10, no.~1, p. 3007, 2019.

\bibitem{Bergholm_2022}
V.~Bergholm, J.~Izaac, M.~Schuld, C.~Gogolin, S.~Ahmed, V.~Ajith, M.~S. Alam, G.~Alonso-Linaje, B.~AkashNarayanan, A.~Asadi \emph{et~al.}, ``Pennylane: Automatic differentiation of hybrid quantum-classical computations,'' \emph{arXiv:1811.04968}, 2022.

\bibitem{Broughton_2021}
M.~Broughton, G.~Verdon, T.~McCourt, A.~J. Martinez, J.~H. Yoo, S.~V. Isakov, P.~Massey, R.~Halavati, M.~Y. Niu, A.~Zlokapa \emph{et~al.}, ``Tensorflow quantum: A software framework for quantum machine learning,'' \emph{arXiv preprint arXiv:2003.02989}, 2021.

\bibitem{Jamadagni_2024}
A.~Jamadagni, A.~M. L{\"a}uchli, and C.~Hempel, ``Benchmarking quantum computer simulation software packages,'' \emph{arXiv:2401.09076}, 2024.

\bibitem{Paszke_2019}
A.~Paszke, S.~Gross, F.~Massa, A.~Lerer, J.~Bradbury, G.~Chanan, T.~Killeen, Z.~Lin, N.~Gimelshein, L.~Antiga \emph{et~al.}, ``Pytorch: An imperative style, high-performance deep learning library,'' \emph{Advances in neural information processing systems}, vol.~32, 2019.

\bibitem{Crooks_2019}
G.~E. Crooks, ``Gradients of parameterized quantum gates using the parameter-shift rule and gate decomposition,'' \emph{arXiv:1905.13311}, 2019.

\bibitem{Wiedmann_2023}
M.~Wiedmann, M.~H{\"o}lle, M.~Periyasamy, N.~Meyer, C.~Ufrecht, D.~D. Scherer, A.~Plinge, and C.~Mutschler, ``An empirical comparison of optimizers for quantum machine learning with spsa-based gradients,'' in \emph{IEEE International Conference on Quantum Computing and Engineering (QCE)}, vol.~1, 2023, pp. 450--456.

\bibitem{Jones_2020}
T.~Jones and J.~Gacon, ``Efficient calculation of gradients in classical simulations of variational quantum algorithms,'' \emph{arXiv:2009.02823}, 2020.

\bibitem{Perez_2020}
A.~P{\'e}rez-Salinas, A.~Cervera-Lierta, E.~Gil-Fuster, and J.~I. Latorre, ``Data re-uploading for a universal quantum classifier,'' \emph{Quantum}, vol.~4, p. 226, 2020.

\bibitem{Jerbi_2021a}
S.~Jerbi, C.~Gyurik, S.~Marshall, H.~Briegel, and V.~Dunjko, ``Parametrized quantum policies for reinforcement learning,'' \emph{Adv. Neural Inf. Process. Syst.}, vol.~34, pp. 28\,362--28\,375, 2021.

\bibitem{MNIST_1998}
\BIBentryALTinterwordspacing
Y.~LeCun, C.~Cortes, and C.~J. Burges, ``The {MNIST} database of handwritten digits,'' 1998. [Online]. Available: \url{http://yann.lecun.com/exdb/mnist/}
\BIBentrySTDinterwordspacing

\bibitem{Periyasamy_2022}
M.~Periyasamy, N.~Meyer, C.~Ufrecht, D.~D. Scherer, A.~Plinge, and C.~Mutschler, ``Incremental data-uploading for full-quantum classification,'' in \emph{IEEE International Conference on Quantum Computing and Engineering (QCE)}, 2022, pp. 31--37.

\bibitem{Meyer_2023a}
N.~Meyer, D.~Scherer, A.~Plinge, C.~Mutschler, and M.~Hartmann, ``Quantum {P}olicy {G}radient {A}lgorithm with {O}ptimized {A}ction {D}ecoding,'' in \emph{International Conference on Machine Learning (ICML)}, vol. 202.\hskip 1em plus 0.5em minus 0.4em\relax PMLR, 2023, pp. 24\,592--24\,613.

\bibitem{Brockman_2016}
G.~Brockman, V.~Cheung, L.~Pettersson, J.~Schneider, J.~Schulman, J.~Tang, and W.~Zaremba, ``{O}pen{AI} {G}ym,'' \emph{arXiv:1606.01540}, 2016.

\bibitem{Periyasamy_2024}
M.~Periyasamy, A.~Plinge, C.~Mutschler, D.~D. Scherer, and W.~Mauerer, ``Guided-spsa: Simultaneous perturbation stochastic approximation assisted by the parameter shift rule,'' \emph{arXiv:2404.15751}, 2024.

\bibitem{Stokes_2020}
J.~Stokes, J.~Izaac, N.~Killoran, and G.~Carleo, ``{Q}uantum {N}atural {G}radient,'' \emph{Quantum}, vol.~4, p. 269, 2020.

\bibitem{Meyer_2023b}
N.~Meyer, D.~D. Scherer, A.~Plinge, C.~Mutschler, and M.~J. Hartmann, ``Quantum natural policy gradients: Towards sample-efficient reinforcement learning,'' in \emph{IEEE International Conference on Quantum Computing and Engineering (QCE)}, vol.~2, 2023, pp. 36--41.

\bibitem{Thanasilp_2023}
S.~Thanasilp, S.~Wang, N.~A. Nghiem, P.~Coles, and M.~Cerezo, ``Subtleties in the trainability of quantum machine learning models,'' \emph{Quantum Mach. Intell.}, vol.~5, no.~1, p.~21, 2023.

\end{thebibliography}

\end{document}